# Reconstruction of the magnetic perturbation in a toroidal force-free circular plasma: application to the Reversed Field Pinch

David Terranova and Paolo Zanca

*Consorzio RFX, Associazione Euratom-ENEA sulla fusione*

**Abstract**
A new method to obtain the radial profile of the magnetic perturbation in a toroidal force-free plasma having circular cross section is developed. The method is quite general and can be applied to any circular low-beta plasma once suitable boundary conditions are imposed. In particular, due to its characteristics, it is appropriate for describing the Reversed Field Pinch (RFP) configuration.

The curvilinear metric of the toroidal geometry is described by adopting flux co-ordinates, and the magnetic field is expressed by the contravariant representation with two independent potentials. The method leads to a system of Newcomb-like equations in the two independent potentials for modes with the same toroidal *n*-number and poloidal mode number *m* coupled by toroidicity.

The advantage of this description is that the potentials together with the metric coefficients can be related in a straightforward manner to the magnetic field. Moreover it is formally simple and maintains a similarity with the cylindrical treatment. Indeed it is customary in RFPs to approach the analysis of MHD instabilities by using cylindrical geometry. Nonetheless the effect of a more realistic toroidal geometry can play an important role, and indeed by applying the method to the Reversed Field eXperiment (RFX) plasma we found that the toroidal effects on the magnetic perturbations are not negligible.

**Introduction and model description**

In a reversed field pinch (RFP) plasma the equilibrium configuration is sustained by means of the dynamo mechanism provided mainly by tearing modes [1,2]. Typically these harmonics are present with a broad toroidal *n*-number spectrum with poloidal mode numbers mostly $m=1$, $m=0$. In RFP plasmas it is customary to study tearing modes assuming cylindrical geometry, nonetheless in the actual experiments the aspect ratio is not so large as to make the toroidal effects negligible. The toroidicity produces poloidal harmonics in the equilibrium quantities (at the leading order $m=\pm 1$, $n=0$), which act as mediators between modes with the *same* toroidal number and *different* poloidal numbers. This kind of interaction is called *toroidal coupling* and plays an important role in tokamak configuration [3,4].

In [5] we presented a new method for mode analysis in *a realistic toroidal geometry for a force-free plasma with circular cross section*. The interested reader can refer to that work for details. The starting point is the contravariant representation of the magnetic field:

1) $\quad \mathbf{B} = \nabla F \times \nabla \vartheta - \nabla \Psi \times \nabla \phi$

where the toroidal and poloidal covariant components of the vector potential are respectively

2) $\quad \Psi(r,\vartheta,\phi) \equiv \Psi_0(r) + \sum_{\substack{n \neq 0 \\ m}} \psi^{m,n}(r) e^{i(m\vartheta - n\phi)}$,

3) $\quad F(r,\vartheta,\phi) \equiv F_0(r) + \sum_{\substack{n \neq 0 \\ m}} f^{m,n}(r) e^{i(m\vartheta - n\phi)}$.

Here flux co-ordinates referred to non-concentric unperturbed circular magnetic surfaces (*r*: magnetic surface radius; $\vartheta$: poloidal-like angle; $\phi$: toroidal angle) are adopted. The quantities $F_0$ and $\Psi_0$ can be identified respectively with the toroidal and poloidal flux of the unperturbed magnetic surfaces divided by $2\pi$. The complex harmonics $f^{m,n}$, $\psi^{m,n}$ represent the perturbations. Through the linearized ideal force balance equation the model leads to a system of Newcomb-like equations in the two independent potentials for the modes with the same *n*-number and *m*-number coupled by toroidicity:

$$4)\ n\left(\frac{g_{\vartheta\vartheta}}{\sqrt{g}}\right)^{0,0}\frac{d\psi^{m,n}}{dr}+mK(r)\frac{df^{m,n}}{dr}-\sigma\left(nf^{m,n}-m\psi^{m,n}\right)-in\left(\frac{g_{r\vartheta}}{\sqrt{g}}\right)^{1,0}\left[nf^{m+1,n}+\right.$$

$$\left.-(m+1)\psi^{m+1,n}-nf^{m-1,n}+(m-1)\psi^{m-1,n}\right]+n\left(\frac{g_{\vartheta\vartheta}}{\sqrt{g}}\right)^{1,0}\left[\frac{d\psi^{m-1,n}}{dr}+\frac{d\psi^{m+1,n}}{dr}\right]=0\ ;$$

$$5)\ -\frac{d}{dr}\left(K(r)\frac{df^{m,n}}{dr}\right)-\left(\frac{g_{rr}}{\sqrt{g}}\right)^{0,0}\left[mn\psi^{m,n}-n^2 f^{m,n}\right]-\sigma\frac{d\psi^{m,n}}{dr}+\frac{nf^{m,n}-m\psi^{m,n}}{m-nq}\frac{d\sigma}{dr}+$$

$$-n\left(\frac{g_{rr}}{\sqrt{g}}\right)^{1,0}\left[(m-1)\psi^{m-1,n}-nf^{m-1,n}+(m+1)\psi^{m+1,n}-nf^{m+1,n}\right]+in\left(\frac{g_{r\vartheta}}{\sqrt{g}}\right)^{1,0}\left[\frac{d\psi^{m+1,n}}{dr}-\frac{d\psi^{m-1,n}}{dr}\right]=0$$

Here $\sigma=\mu_0\mathbf{J}_0\cdot\mathbf{B}_0/B_0^2$. The metric quantities $K$, $g_{ij}/\sqrt{g}$ are computed by the standard perturbative expansion in the parameter $\varepsilon\equiv b/R_0$ (the ratio between the shell minor radius $b$ and major radius $R_0$) [3, 4].

By using this reconstruction it is possible to discriminate between different poloidal harmonics using external magnetic fluctuations measurements. The method is applied to the RFX experiment [6] and considering the specific probes layout (two toroidal arrays of 72 pick-up coils for the toroidal field measurement on the inner surface of the *ideal shell* at opposite poloidal angles $\theta_m$, $\theta_m+\pi$) one has the following limitations on the reconstructed harmonics: $m=-1,0,1,2$ and $n\leq 13$. In this interval, for the typical RFX equilibrium, the $m=2$, $m=-1$ modes are toroidally generated non-resonant sideband of the $m=1$, $m=0$ respectively. The $m=0$ are resonant at the reversal surface and the $m=1$ are resonant for $n\geq 7$. The latter are the dominant dynamo modes of RFX.

For $n>14$ the $m=2$ harmonics become resonant and further measurements, not available at the time of the last experimental campaign of RFX, would be necessary to get the mode reconstruction.

**Application to experimental data**

In RFX the analysis of magnetic fluctuations has been done so far with the assumption of cylindrical geometry [7]. In this context the *half-sum* and *half-difference* of the signals from the two arrays were related respectively to the *even-m* and *odd-m* components of the fluctuation respectively. These two branches were then identified with the $m=0$ harmonics (the *even-m*) and $m=1$ harmonics (the *odd-m*). The new toroidal analysis highlights that, though this identification is fairly adequate for $m=1$ harmonics, it is generally misleading for $m=0$. Indeed the expression of the $n^{th}$ toroidal component of the half-sum $S_n$, emerging from the toroidal reconstruction, is a mixture of various contributions, which can have similar amplitude, not directly distinguishable in the cylindrical analysis[5]:

$$6)\ S_n=\left.\frac{df^{-1,n}}{dr}\right|_b e^{-i\left[\theta_m+\lambda_2(b)\sin(2\theta_m)+\frac{\pi}{2}\right]}\sin[\lambda_1(b)\sin(\theta_m)]+\left.\frac{df^{0,n}}{dr}\right|_b+$$

$$+\left.\frac{df^{1,n}}{dr}\right|_b e^{i\left[\theta_m+\lambda_2(b)\sin(2\theta_m)+\frac{\pi}{2}\right]}\sin[\lambda_1(b)\sin(\theta_m)]+\left.\frac{df^{2,n}}{dr}\right|_b e^{i[2\theta_m+2\lambda_2(b)\sin(2\theta_m)]}\cos[2\lambda_1(b)\sin(\theta_m)],$$

Here $\lambda_1=o(\varepsilon)$, $\lambda_2=o(\varepsilon^2)$ are quantities related to the toroidal geometry. As can be seen $S_n$ gets contributions from all $m=-1,0,1,2$ with different weights. When $n\leq 6$ $S_n$ is related without ambiguity to the $m=0$ poloidal harmonic, because only this mode is resonant while the other harmonics are toroidally induced sideband with negligible contributions in (6). But in correspondence of the dominant ($n\geq 7$) $m=1$ modes, the last three terms related to the $m=0,1,2$ harmonics have comparable amplitude (the $m=-1$ term is always very small). In this case one cannot identify $S_n$ with any particular ($m,n$) harmonic.

In figure (1) we show a comparison between the spectra obtained with the standard cylindrical analysis of the edge data (left), and with the toroidal reconstruction model (right). The harmonics refer to the edge toroidal magnetic field. The reconstructed $m=1$ modes are almost identical to the "cylindrical" $m$-odd for all toroidal numbers. Differently for the $m$-even, since we have shown they are a mixture of the $m=0,1,2$ harmonics. Notice that, while the (0, $n\leq 6$) modes are almost identical to the ($m$-even, $n\leq 6$), the trend of the (0, $n\geq 7$) harmonics departs from the ($m$-even, $n\geq 7$) spectrum, indeed showing a clear contribution from the dominant (1,$n\geq 7$) modes. The same is seen for the (2, $n\geq 7$) modes, because they are totally generated as sidebands of the corresponding $m=1$ modes. Differently from the $m=2$, the shape of the (0, $n\geq 7$) spectrum cannot be explained only in terms of the geometrical $m=0$ sideband from the $m=1$, since we have found that the contribution of this sideband is not significant at the reversal surface. Therefore we can conclude the system is undergoing a dynamical process: the already unstable $m=0$, $n\geq 7$ modes are further destabilised by the toroidal coupling with the corresponding $m=1$.

An indication of the reliability of our analysis is also obtained when looking at spontaneous mode rotations (figure 2). A spontaneous rotation of an ($m$-odd, $n$) mode seen in the measured data is recovered in the reconstructed ($m=1$, $n$) harmonic and is absent from the ($m=0$, $n$) one. At the same time one can clearly see the toroidal generation of $m=2$ mode that follows the rotation of the $m=1$ mode, and is independent from the $m=0$ harmonic. This proves that the various harmonic contributions are correctly identified.

The solution of equations (4, 5) provides complete information on the radial field profile of the modes showing some differences with respect to the cylindrical case. In figure 3 we compare the radial field profile coming from our toroidal analysis with the correspondent profile obtained by solving Newcomb's equation in cylindrical geometry for the mode (0,8). The two functions are quite different especially in the core, where the coupling with the resonant mode (1,8) gives the largest contribution.

The knowledge of the profiles allows in particular the determination of the radial field amplitudes at the respective resonances of each mode. This "radial field spectrum" is useful when comparing magnetic topology and tomographic reconstructions. In RFX we have evidence of the presence of a well defined thermal structure in correspondence with particular regimes where a dominant (1, $n$) mode is present in the magnetic spectrum [8]. This correspondence is clearer when considering, in place of the edge toroidal field spectrum, the radial field spectrum. As an example, in figure 4 we show the spectrum at the edge (left) for the toroidal field fluctuations, and that of the radial field at the resonances (right). The plot refers to a shot that has a thermal structure: the spectrum at the edge is unable to explain the tomographic result, but this can be clearly interpreted from the radial field spectrum.

We have also verified that the shapes of the thermal structure and of the island associated to the dominant mode, as reconstructed by our method, are compatible (figure 5).

**Conclusions**

We have specialized to the RFX experiment a general method for obtaining the radial profile of magnetic field fluctuations in toroidal geometry. Though typically RFP machines are well described in cylindrical geometry, the coupling of magnetic fluctuations due to toroidal geometry cannot be totally neglected. In particular this is a dynamical mechanism that acts as a further destabilization source for the magnetic field perturbation.

We have shown that the interpretation of experimental data in some cases can be ambiguous if considered in cylindrical geometry, but is well interpreted when taking into account the more realistic configuration.

## List of figures

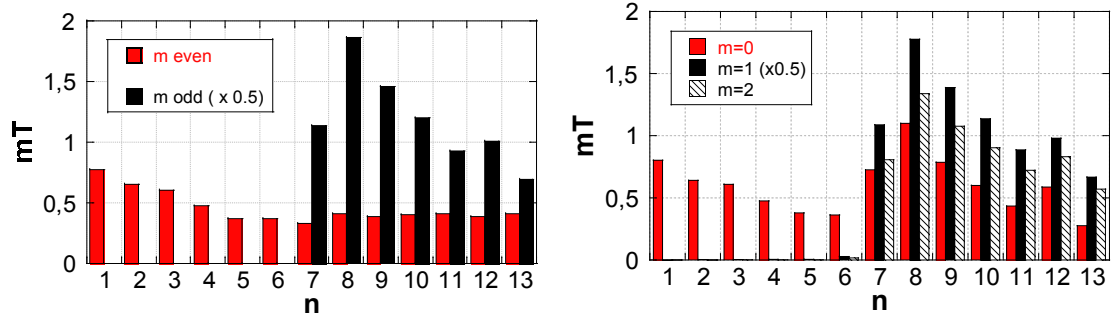

**Figure 1.** Spectral decomposition of the edge toroidal field. Left: measured data adopting cylindrical geometry. Right: results from the toroidal reconstruction model. In the latter case the distinction of the *m*=2 sideband allows a better determination of the *m*=0, *n*≥7 modes.

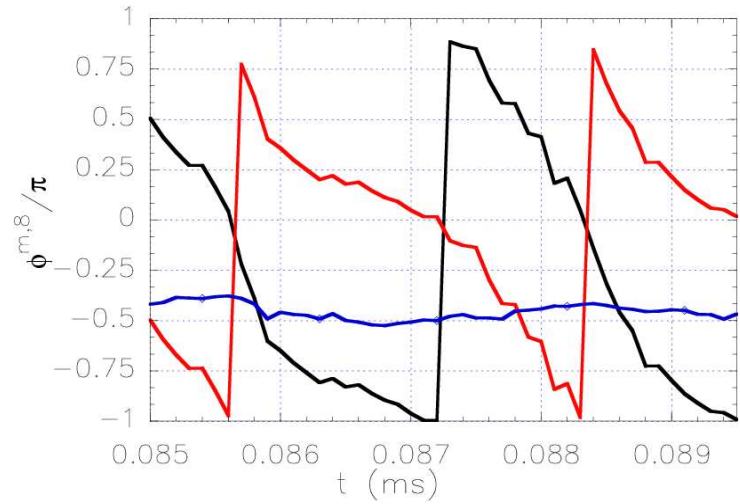

**Figure 2.** Spontaneous rotation of an *m*=1 mode (black): the *m*=2 mode (red) is toroidally generated by the *m*=1 mode and follows its rotation, while the corresponding *m*=0 mode (blue) does not rotate.

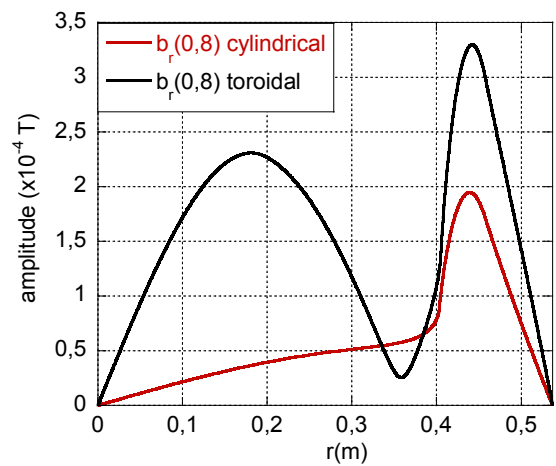

**Figure 3.** The *m*=0, *n*≥7 modes are strongly affected by the toroidal coupling with the *m*=1, *n*≥7 modes: the radial field profile obtained in the toroidal analysis is very different from the cylindrical one.

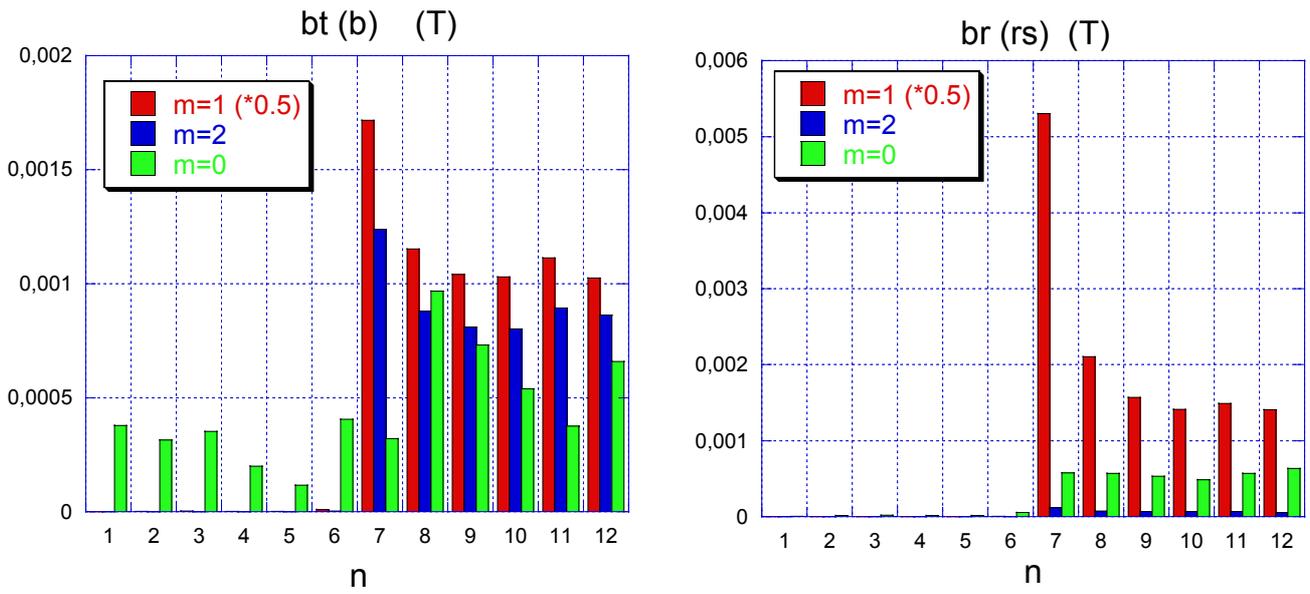

**Figure 4.** Spectrum of magnetic fluctuations for RFX: toroidal component at the edge (left) and radial component at resonances (right). Since the considered $m=2$ modes are not resonant we provide their amplitude at the resonance of the corresponding $m=1$ mode. In the right plot the presence of a dominant $m=1$, $n=7$ mode is evident.

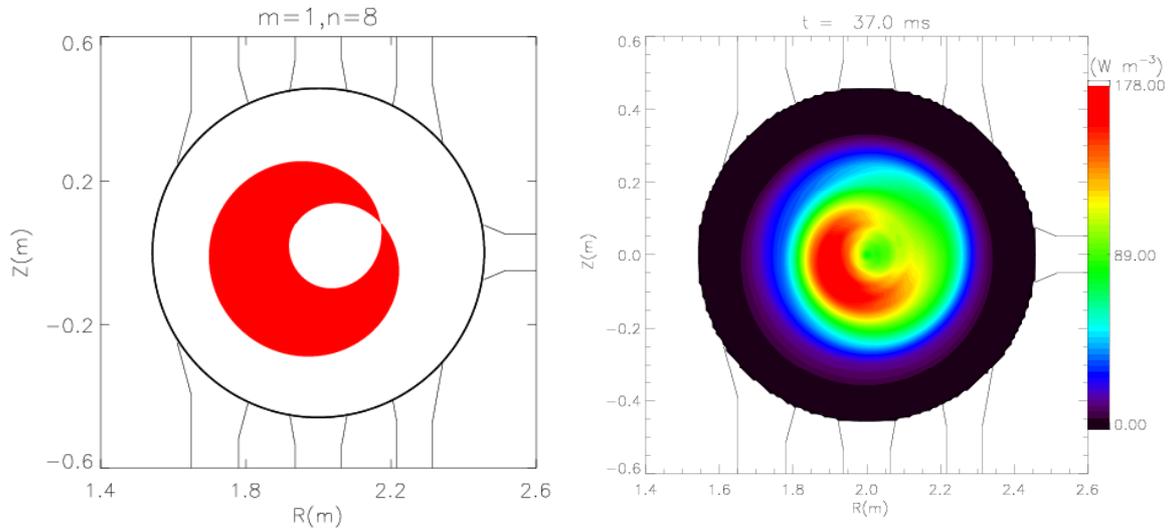

**Figure 5.** Comparison between the toroidal reconstruction of the magnetic island associated to the dominant harmonic (in this shot the $m=1$, $n=8$ mode) and the thermal structure observed with the tomography.